\documentclass[12pt]{article}
\usepackage{lipsum}
\usepackage{amsmath}
\usepackage{amssymb}
\usepackage[normalem]{ulem}
\usepackage{lineno}
\usepackage{longtable}
\usepackage{booktabs}
\usepackage{svg}
\usepackage{adjustbox}
\usepackage{ragged2e}
\usepackage{multirow}
\usepackage{tabularx}
\usepackage{pdfpages}
\usepackage{pdflscape}
\usepackage{makecell}
\usepackage{enumitem}
\usepackage{caption}
\usepackage{booktabs}

\PassOptionsToPackage{hyphens}{url}
\usepackage{hyperref}
\usepackage[vmargin=0.5in]{geometry}

\usepackage[numbers,sort&compress,super]{natbib}

\usepackage{graphicx}
\usepackage{caption}
\usepackage{subcaption}
\makeatother

\renewcommand{\baselinestretch}{1}
\makeatletter
\let\saved@includegraphics\includegraphics
\AtBeginDocument{\let\includegraphics\saved@includegraphics}
\makeatother

\title{Comorbidity Network Analysis Reveals Diagnostic Disparities Between Austrian and Non-Austrian Inpatients: A Population-Wide Cohort Study}

\author{
  Elma Dervić$^{1,2,3}$, 
  Andrea Vismara$^{2}$, 
  Remah Dahdoul$^{1,2}$,\\
  Marion Aichberger$^{4}$, 
  Nilufar Mossaheb$^{4}$
}

\date{}

\begin{document}
\maketitle

\noindent
$^{1}$Medical University of Vienna, Center for Medical Data Science, Institute of the Science of Complex Systems, Spitalgasse 23, 1090 Vienna, Austria\\
$^{2}$Complexity Science Hub, Metternichgasse 8, 1030 Vienna, Austria\\
$^{3}$Supply Chain Intelligence Institute Austria (ASCII), Metternichgasse 8, 1030 Vienna, Austria\\
$^{4}$Department of Psychiatry and Psychotherapy, Clinical Division of Social Psychiatry, Medical University of Vienna, Vienna, Austria\\
\noindent$^{*}$Correspondence: dervic@csh.ac.at

\begin{abstract}
International migrants face well-documented barriers to healthcare access, yet the extent to which these barriers shape patterns of disease co-occurrence remains poorly understood. Drawing on a nationwide dataset of approximately 13 million hospital admissions from around 4 million individuals in Austria (2015–2019), we constructed and compared comorbidity networks between Austrian nationals and non-Austrian migrants, matched 1:1 by age, sex, and time of first hospital admission (272,779 per group).

Following matching, metabolic and cardiovascular diagnoses, including type 2 diabetes and myocardial infarction, were more common among non-Austrians, while depression was more common among Austrians. Comorbidity network analysis showed that among all disease pairs that differed significantly between groups, 70\% showed stronger co-occurrence in Austrian patients and 30\% in non-Austrian patients. Distinct sex-specific patterns appeared: Austrian males showed stronger associations between alcohol use disorder and mental health diagnoses, whereas non-Austrian males more frequently presented with acute somatic conditions. Among non-Austrian women, a pronounced cluster of recurrent depression, somatoform disorders, and dorsalgia was observed.

We interpret the disproportionately fewer comorbidity links observed in non-Austrians not as evidence of lower disease burden, but as a likely reflection of structural access barriers, including language differences, cultural factors, and crisis-oriented admission patterns, that prevent comprehensive diagnostic assessment, though a contribution from the healthy migrant effect cannot be excluded. These findings stress the need for culturally aware care strategies and earlier identification of high-risk multimorbidity profiles in migrant populations.
\end{abstract}

Healthcare $|$ Migration $|$ Comorbidity Networks
\maketitle

\subsection*{Research in context}

\textbf{Evidence before this study}
Prior work has documented lower primary care use and greater reliance on emergency care among migrants, and one Austrian population-wide study found lower hospitalization rates but higher readmission rates among non-nationals. No study had examined the comorbidity network structure in migrants versus natives using nationwide inpatient data.

\textbf{Added value of this study}
This study provides the first population-scale comorbidity network analysis comparing Austrian and non-Austrian inpatients. By moving beyond prevalence comparisons to structured co-occurrence analysis in matched cohorts, we show that non-Austrians have systematically fewer comorbidity links, a finding that persists across ICD-10 chapters and both sexes and may be better explained by access barriers rather than lower disease burden.

\textbf{Implications of all the available evidence}
Healthcare systems in high-migration countries should invest in culturally sensitive diagnostic pathways and earlier identification of multimorbidity in migrant populations. The comorbidity patterns identified here, particularly musculoskeletal–mental health clusters in non-Austrian women and acute alcohol-related presentations in non-Austrian men, can direct targeted care navigation procedures.

\section*{Introduction}


International migrants tend to have poorer access to healthcare services than natives. In some cases, this is due to the institutional design of national healthcare systems \cite{who2022refugees}. In other cases, inequalities arise from socio-cultural factors within migrant communities, including language and communication barriers, as well as patterns of healthcare utilization such as lower use of primary care and potential reliance on emergency services \cite{lebano2020migrants}. Inequality and discrimination in healthcare operate through multiple mechanisms, including stereotyping and the unequal distribution of diagnostic and therapeutic resources, with migrants and minorities disproportionately affected \cite{hamed2022racism, pattillo2023racism}. Another contributing factor is the distinct pattern of healthcare utilization among migrants: refugees and recent arrivals tend to rely less on primary care initially, but this pattern often changes over time. Evidence on emergency care utilization remains mixed, with some studies suggesting overuse and others finding no such effect \cite{topal2012challenges, acquadro2024exploring}. These dynamics shape how migrants engage with health services, contributing to lower adherence to treatment regimes \cite{christian2006quality} and reduced participation in preventive care \cite{benkert2006effects, burgess2008association}. This reality persists across EU countries \cite{gil2021access}.

The consequences of these access barriers are substantial. Limited access to healthcare negatively affects migrants' quality of life \cite{norredam2011migrants}. For example, language barriers between healthcare staff and patients are associated with reduced symptom reporting and fewer referrals to secondary and specialist care \cite{bischoff2003language}. Over time, unmet healthcare needs can lead to a worsening burden of disease, which not only affects individuals but also places additional pressure on healthcare system resources \cite{sleeman2019escalating}. Addressing these challenges requires a better understanding of the conditions faced by international migrants within host countries' healthcare systems, as this can improve both individual well-being and the overall efficiency of healthcare provision.

Austria is a relevant case study, as its healthcare system reproduces some of the challenges in access for migrants described above. A recent study showed that access to healthcare for international migrants in Austria is poor, characterized by lower hospitalisation rates than those of natives but higher readmission rates \cite{dervic2024healthcare}. These patterns, however, should be interpreted cautiously, as lower utilisation may partly reflect the healthy migrant effect rather than structural barriers to access alone. Moreover, Austria is among the EU Member States with the highest shares of foreign-born residents: 22$.$1\% of Austria’s resident population was foreign-born on 1 January 2024, compared with 13$.$t9\% in the EU overall \cite{eurostat_popdiv_2025}. Another study showed how Syrian, Iraqi, and Afghan refugees living in Austria rate their own health as significantly worse than that of natives, as well as documenting the many barriers they face in having access to healthcare \cite{kohlenberger2019barriers}. 

Beyond documenting inequalities in access, important questions remain about how these disparities translate into health outcomes over time. This study addresses this gap by focusing on the comorbidity networks of migrants living in Austria, that is, how different diseases co-occur among the same patients. Pattern of comorbidity are often derived or exacerbated by limited access to healthcare \cite{chudasama2021clustering, azubuike2024delayed}. Moreover, analyzing comorbidities offers a more comprehensive perspective than single-disease approaches, capturing how health conditions evolve and cluster under constrained access to care. This is particularly relevant for understanding healthcare system pressures, as patients with multiple co-occurring conditions tend to require more intensive and repeated care. This study analyses the patterns of comorbidity in Austrian native and migrant patients by using a database of hospitalisation information of all Austrian hospitals between 2015 and 2019. This unique dataset enables comparison of disease development and comorbidities among native and migrant populations to uncover patterns that could help improve healthcare administration in Austria.

Our results reveal substantial differences in disease co-occurrence patterns between Austrian nationals and non-Austrian migrants. In the unmatched cohort, Austrian patients were older, had more hospitalisations, longer stays, and a higher burden of chronic conditions, including hypertension, heart failure, and depression. After age- and sex-matching, metabolic and cardiovascular diagnoses became more prevalent among non-Austrians, including type 2 diabetes and myocardial infarction, while depression remained more frequent among Austrians. Comorbidity network analysis in the matched cohort showed that significant comorbidity differences were predominantly found in the Austrian network (70\% vs. 30\% in non-Austrians), suggesting that non-Austrians receive less comprehensive diagnostic assessment, likely reflecting access barriers such as language differences, cultural factors, and more crisis-oriented admission patterns. Sex-specific patterns were also observed: among Austrian males, comorbidity links within mental health diagnoses (notably alcohol use disorder and depression) were pronounced, whereas non-Austrian males showed a higher prevalence of acute presentations such as alcohol withdrawal-related epileptic seizures. Among non-Austrian women, comorbidities between recurrent depression, somatoform disorders, and musculoskeletal pain were particularly prominent, potentially reflecting the cumulative toll of migration-related social risk factors and occupational exposures.

\section*{Data and Methods}
\subsubsection*{Data}
The dataset includes around 13 million hospital admissions from 4 million individuals in Austria between 2015 and 2019. Patients are 46\% male and 54\% female. Each record contains a unique patient ID, sex, age (in 5-year bands), primary and secondary diagnoses, admission/discharge dates, discharge type, regions of residence and treatment, and nationality \cite{haug2020high, dervic2024unraveling, dervic2024healthcare}. Primary diagnoses indicate the main reason for admission; secondary diagnoses reflect additional conditions.

Diagnoses follow level-4 ICD-10 coding, allowing consistent classification across 12,040 distinct conditions \cite{who}. In this project we use diagnoses level 3. The average age is 52$.$8 years (SD = 24$.$4), with Austrian nationals averaging 53$.$9 (SD = 24$.$4) and non-Austrians 40 (SD = 18). Patients with multiple nationalities (1$.$63\%) were excluded. Non-Austrian nationals make up 9$.$4\% of the cohort (N = 327,394), with 57$.$9\% female and 42$.$1\% male \cite{dervic2024healthcare}.

\subsubsection*{Comorbidity Network Construction}

Comorbidity, or disease-disease networks, represent statistically significant associations between diagnoses, where nodes represent diseases and edges reflect correlations derived from large-scale longitudinal health data.
To quantify disease associations, we use the Odds Ratio (OR), which measures how strongly the presence of one disease is associated with the presence of another. Specifically, the OR compares the odds of observing disease B among patients with disease A to the odds of observing disease B among patients without disease A. An OR greater than 1 indicates that the two diseases co-occur more frequently than expected by chance, while an OR equal to 1 suggests no association. For more accurate estimates that adjust for confounding factors like age, sex, and time, we apply the Cochran–Mantel–Haenszel (CMH) method, which computes a weighted average of ORs across stratified subgroups \cite{kuritz1988general}.
For each pair of ICD-10 3-digit diagnoses (1,080 in total), we generated contingency tables, split across age, sex, and 2-year time intervals (2015–2016 and 2017–2018), resulting in 32 strata (8 age groups × 2 time frames × 2 sexes). To avoid unreliable links driven by large sample sizes, we include only disease pairs with $OR > 1.5$, $p < 0.05$, and at least 100 co-occurring patients \cite{dervic2025comorbidity}. 

\subsubsection*{Statistical Test for Differences Between Austrians' and Non-Austrians' Comorbidities}

To assess differences in comorbidities between Austrian and non-Austrian patients, we computed a standardized difference ($D$) for each link in the comorbidity networks, based on the difference of log odds ratios (log ORs) expressed in units of pooled standard errors:

\[
\text{D} = \frac{\log(\text{OR}_\text{AT}) - \log(\text{OR}_\text{non-AT})}{\sqrt{\text{SE}_\text{AT}^2 + \text{SE}_\text{non-AT}^2}}
\]

Here, $\text{OR}_\text{AT}$ and $\text{OR}_\text{non-AT}$ are the odds ratios for Austrians and non-Austrians, respectively, and $\text{SE}_\text{AT}$ and $\text{SE}_\text{non-AT}$ are their corresponding standard errors. This calculation was performed separately for males and females. To evaluate significance, we tested the null hypothesis that D values are drawn from a normal distribution with mean zero, resulting in a corresponding p-value $P_\text{D}$.

We defined five levels of significance based on $D$ magnitude and corresponding $P_\text{D}$ values:

\begin{itemize}
    \item \textbf{Not significant:} D $\leq$ 2  \hfill ($P_\text{D} \geq 0.045$)
    \item \textbf{Weak:} $2 < \text{D} \leq 3$ \hfill ($0.003 \leq P_\text{D} < 0.045$)
    \item \textbf{Substantial:} $3 < \text{D} \leq 4$ \hfill ($0.00006 \leq P_\text{D} < 0.003$)
    \item \textbf{Strong:} $4 < \text{D} \leq 5$ \hfill ($0.00001 \leq P_\text{D} < 0.00006$)
    \item \textbf{Very strong:} D $> 5$ \hfill ($P_\text{D} < 0.00001$)
\end{itemize}

\section*{Results}

\subsubsection*{General Cohort}
Analysis of Austrian inpatient records from 2015 to 2019 revealed that Austrian hospital patients were older than non-Austrian patients (mean age 53$.$3 vs 39$.$8 years). Austrians had more hospitalisations (2.87 vs 2.09), longer cumulative hospital stays (16$.$1 vs 8$.$8 days), and more diagnoses (2$.$24 vs 1$.$67), Table \ref{tab:baseline}. Chronic conditions were notably more prevalent in Austrians within the unmatched cohort. These include hypertension (I10: 21$.$76\% vs 9$.$38\%), heart failure (I50: 4$.$30\% vs 1$.$07\%), stroke (I63: 2$.$27\% vs 0$.$71\%), and depression (F32: 3$.$21\% vs 1$.$78\%).

\begin{figure*}[!h]
\centering
{\includegraphics[width=0.99\linewidth]{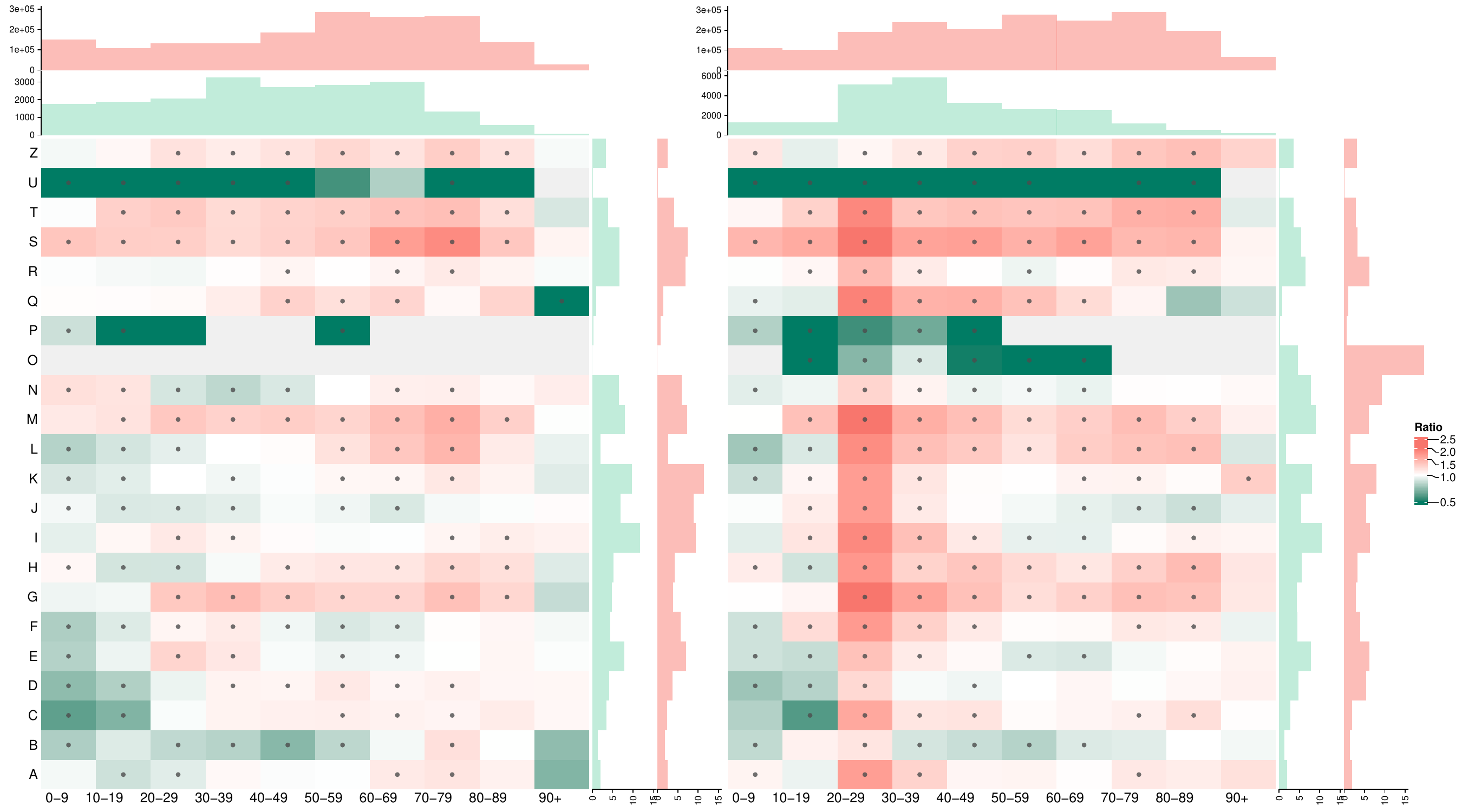}}
\caption{Log-scale prevalence ratios across ICD-10 chapters comparing non-Austrian and Austrian patients, stratified by sex (females: right; males: left). Colored squares represent prevalence differences, with green indicating higher prevalence among non-Austrians and pink among Austrians; grey dots mark statistically significant differences ($p < 0.05$). Grey squares denote cohorts too small to estimate prevalence ratios. Marginal histograms display patient distributions by age group (top) and ICD-10 chapter (right; note: scales differ).}
\label{fig:RatioICDs}
\end{figure*}

\begin{figure*}[!h]
\centering
{\includegraphics[width=0.90\linewidth]{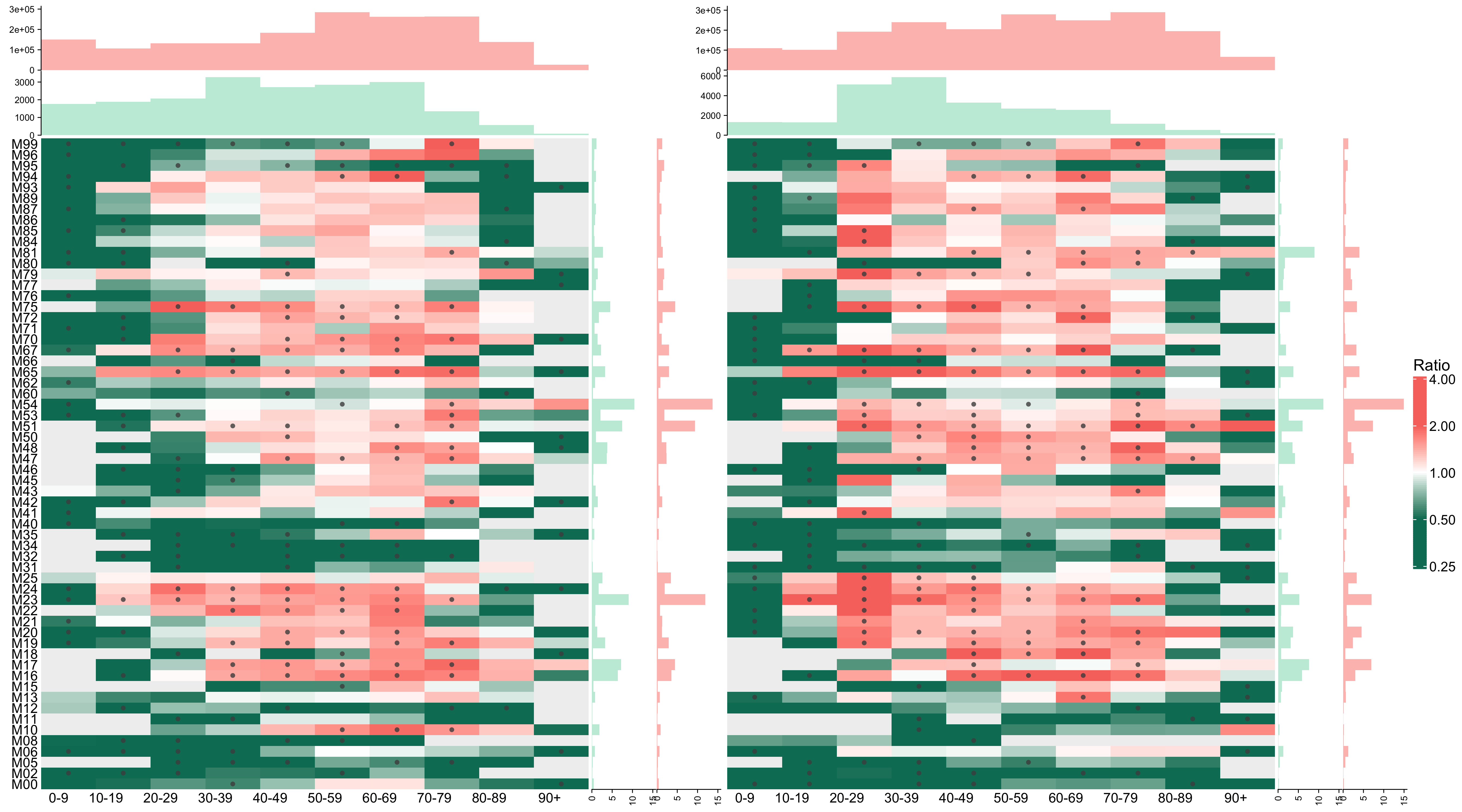}}
\caption{Log-scale ratios of disease prevalence across ICD-10 chapter \textbf{M} - Diseases of the musculoskeletal system and connective tissue - comparing non-Austrian and Austrian patients, stratified by sex (females: right; males: left). Colored squares represent prevalence differences, with green indicating higher prevalence among non-Austrians and pink among Austrians; grey dots mark statistically significant differences ($p < 0.05$). Grey squares denote cohorts too small to estimate prevalence ratios. Marginal histograms display patient distributions by age group (top) and ICD-10 chapter (right; note: scales differ).}
\label{fig:RatioICDsM}
\end{figure*}

Further exploring age-related differences, heatmaps of log-scale prevalence ratios by age group revealed broad, age-dependent differences between Austrians and non-Austrians across ICD-10 chapters. Many chapter–age combinations show statistically significant differences between Austrian and non-Austrian patients (indicated by grey dots), Figure \ref{fig:RatioICDs}. Focusing on the musculoskeletal system and connective tissue diseases (chapter M), differences were observed across numerous 3-digit ICD diagnoses, Figure \ref{fig:RatioICDsM}. This shows that variation is not limited to a few conditions but rather spans many age-specific diagnoses. 

\begin{table*}[ht]
\caption{General cohort baseline characteristics and selected ICD-10 codes by group}
\label{tab:baseline}
\centering
\scriptsize
\renewcommand{\arraystretch}{1} 
\setlength{\tabcolsep}{3pt} 
\begin{tabular}{c|lll|lll}
& \multicolumn{3}{c|}{\textbf{Austrians}} 
 & \multicolumn{3}{c}{\textbf{Non-Austrians}} \\
 & All & Female & Male & All & Female & Male  \\ 
    \hline
        \hline
        All patients & 3,607,058 & 1,927,211 & 1,679,847 & 327,394 & 189,653 & 137,741 \\
Age & 53$.$32$\pm$24$.$18 & 53$.$98$\pm$24$.$06 & 52$.$56$\pm$24$.$3 & 39$.$81$\pm$18$.$84 & 39$.$33$\pm$17$.$34 & 40$.$45$\pm$20$.$72 \\
Hospital\_stay & 2$.$87$\pm$4$.$16 & 2$.$83$\pm$4$.$02 & 2$.$91$\pm$4$.$33 & 2$.$09$\pm$3$.$34 & 2$.$06$\pm$2$.$94 & 2$.$14$\pm$3$.$81 \\
Hospital\_days & 16$.$07$\pm$31$.$56 & 15$.$89$\pm$31$.$31 & 16$.$29$\pm$31$.$85 & 8$.$81$\pm$21$.$83 & 8$.$27$\pm$19$.$36 & 9$.$56$\pm$24$.$81 \\
Number\_diagnoses & 2$.$24$\pm$1$.$92 & 2$.$22$\pm$1$.$88 & 2$.$25$\pm$1$.$97 & 1$.$67$\pm$1$.$25 & 1$.$69$\pm$1$.$21 & 1$.$65$\pm$1$.$31 \\

E11 & 6$.$57 & 5$.$54 & 7$.$76 & 3$.$86 & 2$.$86 & 5$.$23 \\
E66 & 3$.$82 & 3$.$7 & 3$.$96 & 3$.$19 & 3$.$24 & 3$.$12 \\
F32 & 3$.$21 & 4 & 2$.$3 & 1$.$78 & 1$.$92 & 1$.$59 \\
F33 & 1$.$6 & 1$.$95 & 1$.$21 & 0$.$96 & 0$.$99 & 0$.$93 \\
I10 & 21$.$76 & 20$.$83 & 22$.$83 & 9$.$38 & 7$.$43 & 12$.$06 \\
I21 & 1$.$83 & 1$.$25 & 2$.$49 & 0$.$97 & 0$.$4 & 1$.$76 \\
I50 & 4$.$3 & 4$.$11 & 4$.$53 & 1$.$07 & 0$.$82 & 1$.$42 \\
I63 & 2$.$27 & 2$.$02 & 2$.$55 & 0$.$71 & 0$.$47 & 1$.$05 \\
J18 & 3$.$72 & 3$.$16 & 4$.$37 & 1$.$73 & 1$.$24 & 2$.$41 \\
C00-C99 & 9$.$14 & 8$.$03 & 10$.$41 & 4$.$31 & 3$.$7 & 5$.$16 \\ 
E00-E99 & 22$.$92 & 22$.$65 & 23$.$23 & 13$.$3 & 12$.$06 & 15 \\ 
F00-F99 & 13$.$17 & 13$.$17 & 13$.$18 & 9$.$55 & 7$.$68 & 12$.$13 \\ 
I00-I99 & 32$.$32 & 30$.$31 & 34$.$63 & 15$.$56 & 12$.$24 & 20$.$13 \\ 
O00-O99 & 7$.$22 & 13$.$51 & 0 & 22$.$21 & 38$.$34 & 0 \\ 
   \hline
\hline
\end{tabular}
\end{table*}

\begin{figure*}[!h]
\centering
{\includegraphics[width=0.90\linewidth]{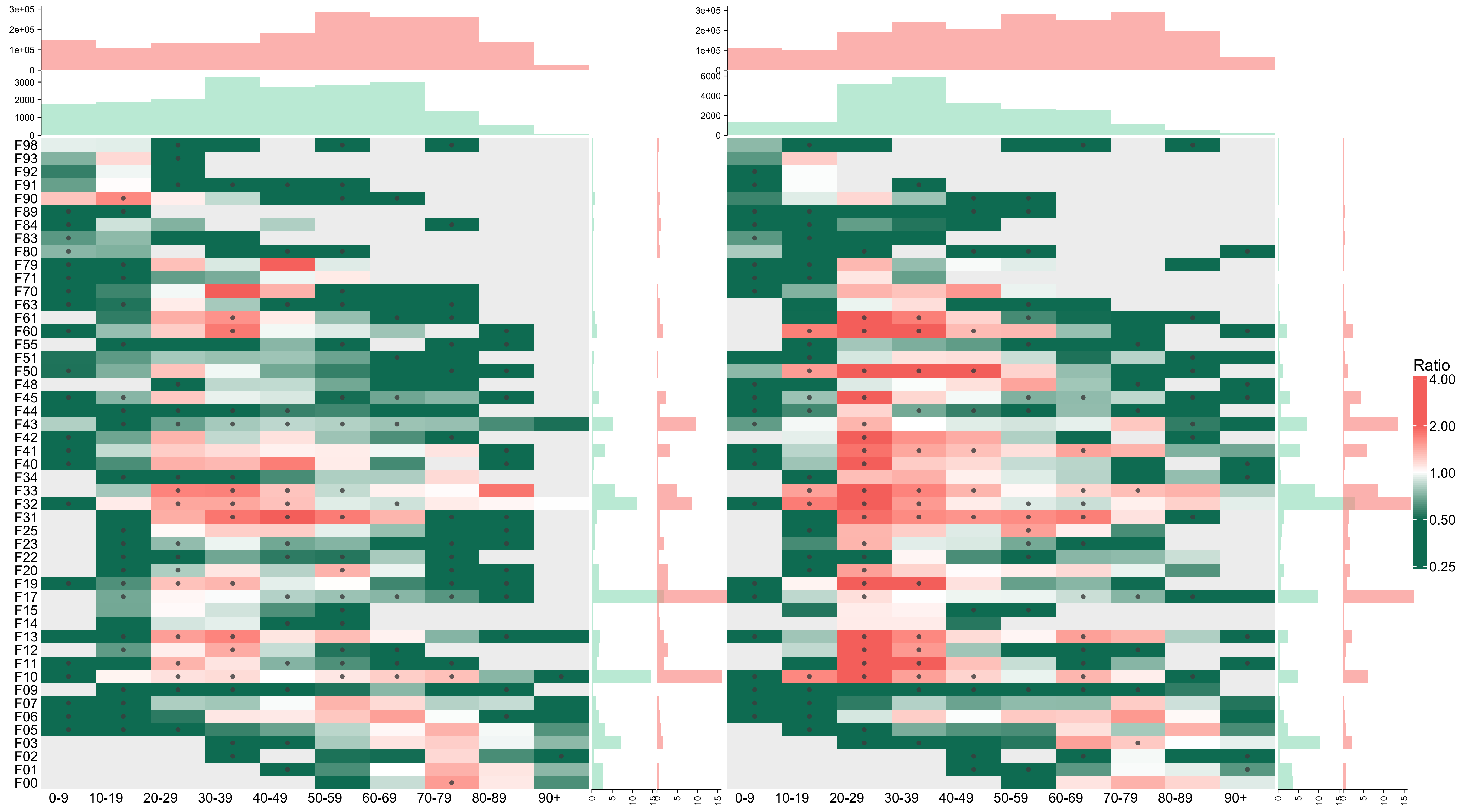}}
\caption{Log-scale ratios of disease prevalence across ICD-10 chapter \textbf{F} - Mental \& behavioural disorders - comparing non-Austrian and Austrian patients, stratified by sex (females: right; males: left). Colored squares represent prevalence differences, with green indicating higher prevalence among non-Austrians and pink among Austrians; grey dots mark statistically significant differences ($p < 0.05$). Grey squares denote cohorts too small to estimate prevalence ratios. Marginal histograms display patient distributions by age group (top) and ICD-10 chapter (right; note: scales differ).}
\label{fig:RatioICDsF}
\end{figure*}

\begin{figure*}[!h]
\centering
{\includegraphics[width=0.90\linewidth]{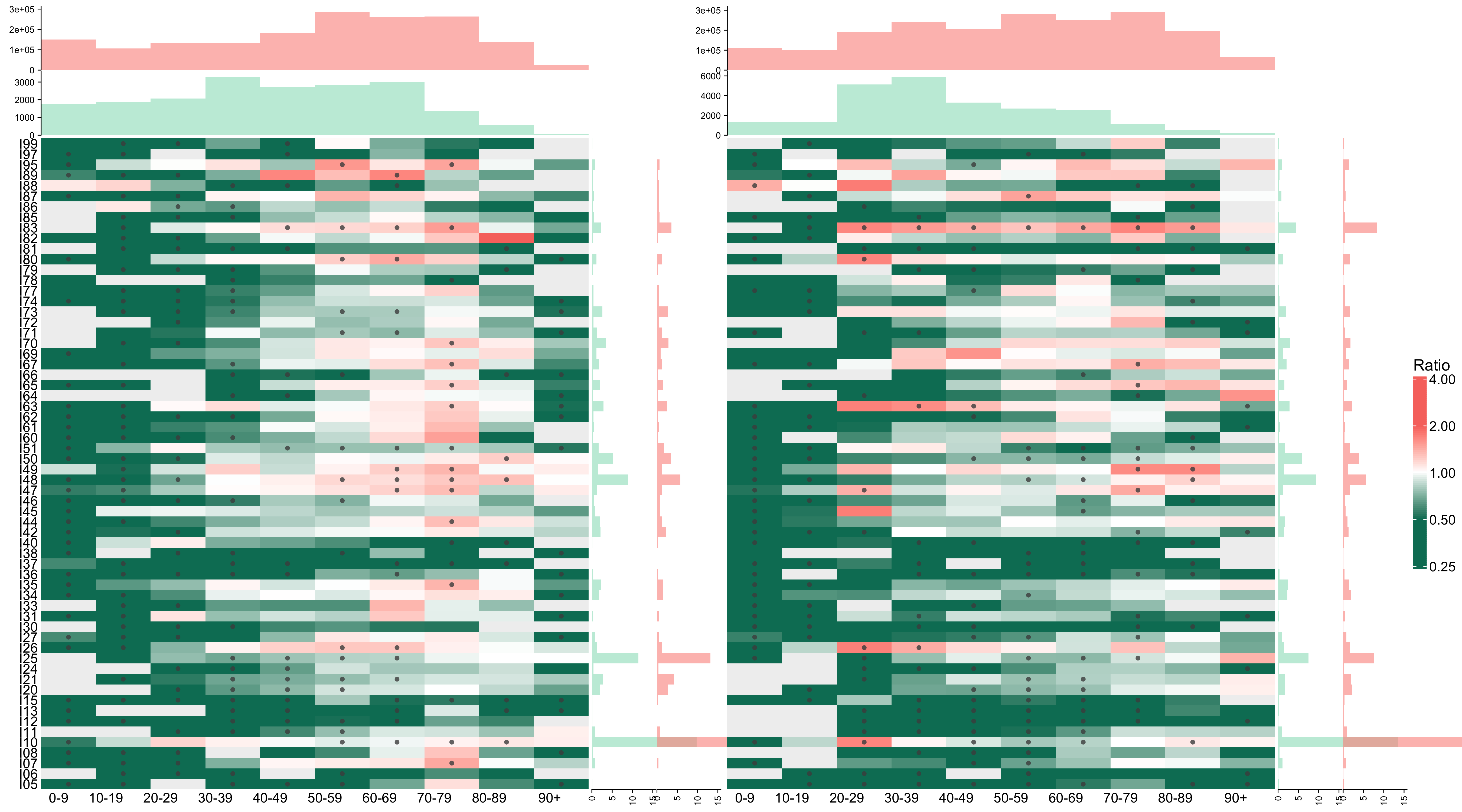}}
\caption{Log-scale ratios of disease prevalence across ICD-10 chapter \textbf{I} - Circulatory system diseases - comparing non-Austrian and Austrian patients, stratified by sex (females: right; males: left). Colored squares represent prevalence differences, with green indicating higher prevalence among non-Austrians and pink among Austrians; grey dots mark statistically significant differences ($p < 0.05$). Grey squares denote cohorts too small to estimate prevalence ratios. Marginal histograms display patient distributions by age group (top) and ICD-10 chapter (right; note: scales differ).}
\label{fig:RatioICDsI}
\end{figure*}

To enable direct comparison, Austrians and non-Austrians were matched 1:1 by age group, sex, and month and year of first hospital visit. This resulted in 272,779 individuals per group. After matching, age distributions were closely aligned (mean age ~40 years in both groups), Table \ref{tab:baselinematched}. Differences in healthcare utilization and diagnosis burden diminished but persisted. Austrians continued to have slightly higher hospital use (hospital days: 11$.$35 vs 9$.$89; diagnoses per stay: 1$.$91 vs 1$.$77). In the matched sample, metabolic and cardiovascular diagnoses became more frequent among non-Austrians. These included type 2 diabetes (E11: 4$.$18\% vs 2$.$84\%), hypertension (I10: 10$.$05\% vs 9$.$38\%), and myocardial infarction (I21: 1$.$03\% vs 0$.$85\%). Depression remained more common among Austrians (F32: 2$.$32\% vs 1$.$97\%; F33: 1$.$46\% vs 1$.$06\%).

\begin{table*}[ht]
\caption{Matched cohort baseline characteristics and selected ICD-10 codes by group}
\label{tab:baselinematched}
\centering
\scriptsize
\renewcommand{\arraystretch}{1} 
\setlength{\tabcolsep}{3pt} 
\begin{tabular}{c|lll|lll}
& \multicolumn{3}{c|}{\textbf{Matched Austrians}} 
 & \multicolumn{3}{c}{\textbf{Non-Austrians}} \\
 & All & Female & Male & All & Female & Male  \\ 
    \hline
        \hline
        All patients & 272,779 & 159,280 & 113,499 & 272,779 & 159,280 & 113,499 \\
Age & 40$.$04$\pm$19$.$06 & 39$.$52$\pm$17$.$52 & 40$.$78$\pm$21 & 39$.$94$\pm$18$.$85 & 39$.$43$\pm$17$.$31 & 40$.$66$\pm$20$.$81 \\
Hospital stay & 2$.$38$\pm$3$.$49 & 2$.$35$\pm$3$.$38 & 2$.$43$\pm$3$.$63 & 2$.$26$\pm$3$.$6 & 2$.$21$\pm$3$.$16 & 2$.$32$\pm$4$.$14 \\
Hospital days & 11$.$35$\pm$27$.$3 & 10$.$8$\pm$26$.$7 & 12$.$11$\pm$28$.$11 & 9$.$89$\pm$23$.$42 & 9$.$23$\pm$20$.$7 & 10$.$81$\pm$26$.$75 \\
Number diagnoses & 1$.$91$\pm$1$.$52 & 1$.$91$\pm$1$.$46 & 1$.$92$\pm$1$.$6 & 1$.$77$\pm$1$.$33 & 1$.$78$\pm$1$.$28 & 1$.$76$\pm$1$.$4 \\

E11 & 2$.$84 & 1$.$8 & 4$.$29 & 4$.$18 & 3$.$09 & 5$.$71 \\
E66 & 3$.$2 & 2$.$98 & 3$.$52 & 3$.$45 & 3$.$52 & 3$.$34 \\
F32 & 2$.$32 & 2$.$57 & 1$.$96 & 1$.$97 & 2$.$12 & 1$.$76 \\
F33 & 1$.$46 & 1$.$59 & 1$.$27 & 1$.$06 & 1$.$09 & 1$.$03 \\
I10 & 9$.$38 & 7$.$03 & 12$.$68 & 10$.$05 & 7$.$92 & 13$.$03 \\
I21 & 0$.$85 & 0$.$37 & 1$.$52 & 1$.$03 & 0$.$43 & 1$.$88 \\
I50 & 1$.$17 & 0$.$79 & 1$.$71 & 1$.$18 & 0$.$91 & 1$.$57 \\
I63 & 0$.$87 & 0$.$61 & 1$.$24 & 0$.$76 & 0$.$51 & 1$.$12 \\
J18 & 1$.$68 & 1$.$22 & 2$.$32 & 1$.$9 & 1$.$35 & 2$.$67 \\
C00-C99 & 4$.$92 & 4$.$18 & 5$.$96 & 4$.$5 & 3$.$81 & 5$.$48 \\ 
E00-E99 & 13$.$89 & 12$.$57 & 15$.$76 & 14$.$11 & 12$.$71 & 16$.$06 \\ 
F00-F99 & 10$.$66 & 9$.$54 & 12$.$22 & 10$.$17 & 8$.$19 & 12$.$96 \\ 
I00-I99 & 16$.$88 & 13$.$5 & 21$.$62 & 16$.$4 & 12$.$9 & 21$.$31 \\ 
O00-O99 & 17$.$64 & 30$.$21 & 0 & 22$.$96 & 39$.$32 & 0 \\ 
\hline
\end{tabular}
\end{table*}

The previous population-wide study showed that non-Austrians utilize the hospital system differently from natives, both in the primary reasons for admission and in the departments they visit. Germans were a notable exception. Simple prevalence comparisons can be misleading because they conflate underlying disease burden with diagnosis and utilization patterns that may vary with healthcare-seeking behavior and access. We therefore matched Austrians and non-Austrians on key covariates and examined comorbidity patterns within these balanced cohorts. This shifts the focus from how often diagnoses are recorded to how conditions co-occur among patients with similar observed characteristics.

\subsubsection*{Comorbidity Network}

Comorbidity networks are structured representations of disease co-occurrence, with nodes representing diagnoses and edges representing the statistically significant association between them. They provide a framework for identifying, analyzing, and comparing comorbidity patterns throughout cohorts. Using this approach, we constructed matched cohorts of Austrians and non-Austrians (matched by age, sex, and month of first hospital visit) and quantified differences in co-occurrence patterns, weighting edges by the standardized difference in log odds ratios between groups.

Overall, the majority of statistically significant differences were observed in the Austrian cohort, accounting for 70\% of them compared with 30\% in the non-Austrian network. Sex-specific patterns also emerged: among Austrians, significant links were nearly balanced between females (49$.$3\%) and males (50$.$7\%), whereas among non-Austrians, they were more male-skewed (62$.$9\% male, 37$.$1\% female). Chapter-level analysis showed that these differences spanned multiple ICD-10 categories for both sexes and were not confined to specific disease groups. All comorbidities with significant differences are listed in Supplementary Tables 1 and 2.

\begin{figure*}[!h]
\centering
{\includegraphics[width=0.99\linewidth]{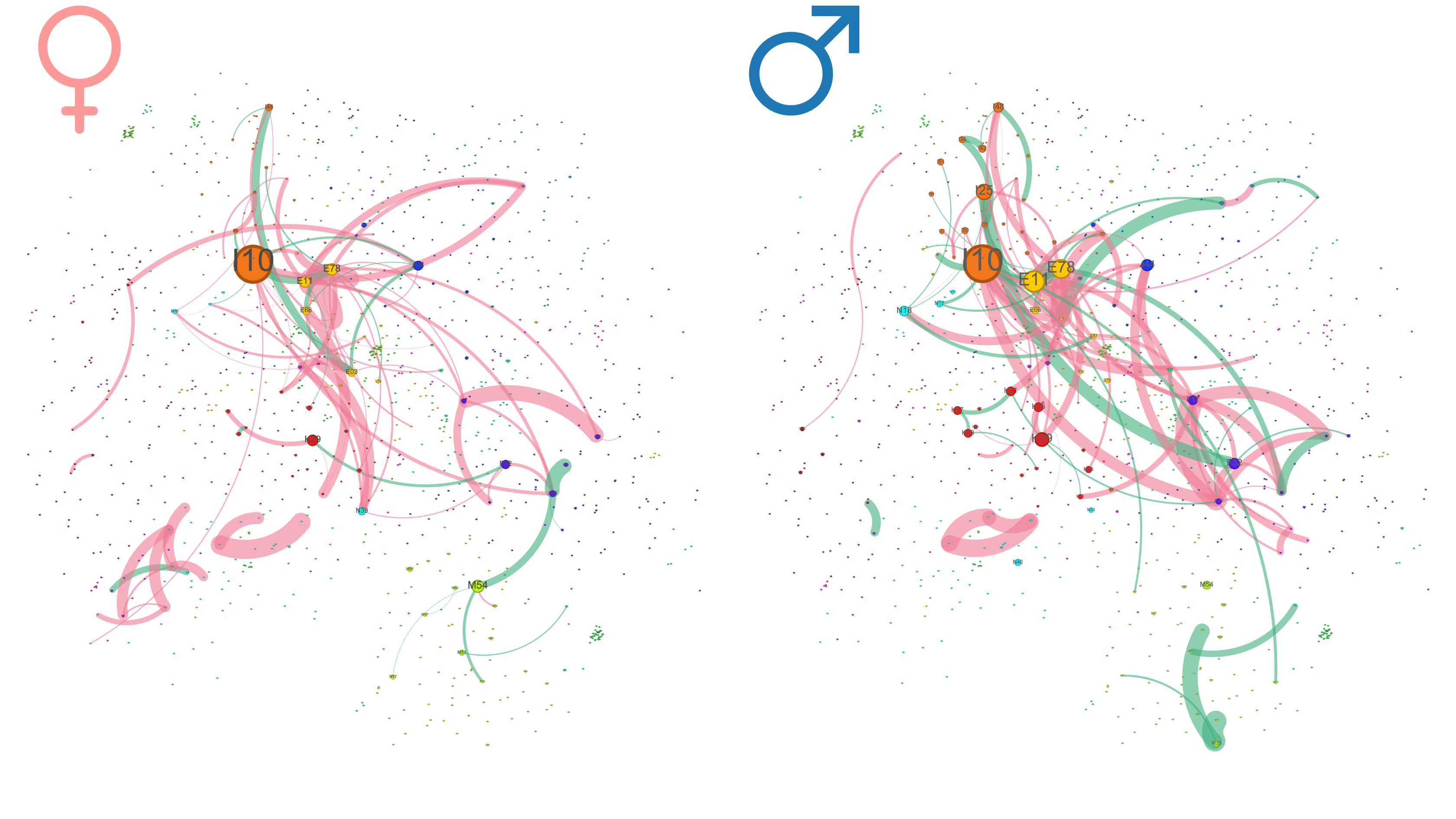}}

\caption{Comorbidity networks highlighting diagnoses that differ significantly between Austrian and non-Austrian patients. Nodes represent ICD-10 3-digit diagnoses; their size reflects node degree, and color denotes the corresponding ICD chapter (based on the first ICD-10 letter). Links represent comorbidities with statistically significant differences between the two groups. Link weights are proportional to the difference in log odds ratios (ORs), scaled by the pooled standard error. Link color indicated the direction of the difference, with pink links denoting stronger co-occurrence among Austrians and green links denoting stronger co-occurrence among non-Austrians. An interactive version is 
available at \url{https://vis.csh.ac.at/diaspora_model_for_migration/migration-net/}} 
\label{fig:RatioComNet}
\end{figure*}

Among females, the largest differences in comorbidity were observed for recurrent depression (F33) with somatoform disorders (F45; standardized difference of log odds ratios - $D = 5.63$: very strong; see Methods) and with dorsalgia (M54; $D = 3.71$), both substantially more frequent in non-Austrian than in matched Austrian women. Among males, the largest comorbidity differences were in musculoskeletal diagnoses, with knee joint disorders (M23–M24) showing the biggest difference overall ($D = 6.39$), alongside a markedly larger difference in the link between depression (F32) and chronic ischemic heart disease (I25; $D = 4.58$).

In Austrian males, the largest differences involving alcohol use disorder (F10) were with nicotine dependence (F17; $D =-6.53$), depression (F32; $D =-3.52$), and adjustment disorder (F43; $D =-3.60$). In non-Austrian males, the largest alcohol-related differences instead pointed to acute somatic presentations, including epileptic seizures (G40; $D = 2.23$) and gastro-oesophageal reflux (K21; $D = 2.33$).

Diabetes-related comorbidities with the largest differences were more often found in Austrians than in non-Austrians in both sexes. The co-occurrence of type 2 diabetes and diabetic retinopathy (E11–H36) showed large differences in both females ($D =-6.54$) and males ($D =-4.02$).

\begin{figure*}[!ht]
\centering
{\includegraphics[width=0.99\linewidth]{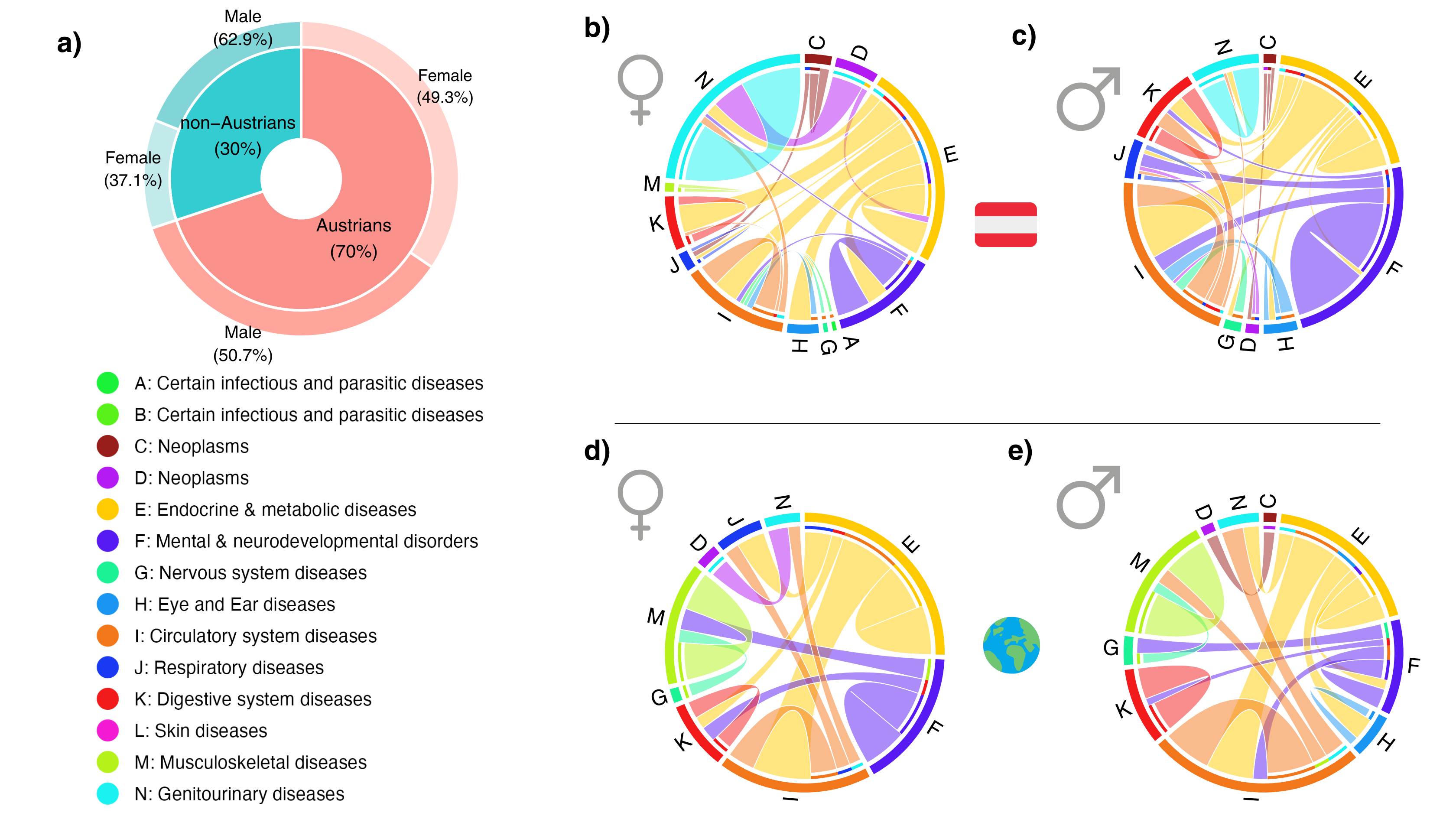}}
\caption{Distribution of significant comorbidity differences a) nationality b) across ICD-10 chapters for males Austrians, c) females Austrians, d) males non-Austrians e) females non-Austrians} 
\label{fig:RatioComNet}
\end{figure*}

\begin{table*}[!h]
\centering
\scriptsize
\caption{Distribution of significant comorbidity differences across ICD-10 chapters by gender and nationality}
\label{tab:icd_links}

\resizebox{\textwidth}{!}{%
\footnotesize
\begin{tabular}{lcccc}
\toprule
 & \multicolumn{2}{c}{Austrians} & \multicolumn{2}{c}{Non-Austrians} \\
\cmidrule(lr){2-3} \cmidrule(lr){4-5}
Chapter & Female & Male & Female & Male \\
\midrule
\textbf{A} Infectious \& Parasitic Diseases & 2 & 2 & 0 & 1 \\
\textbf{B} Infectious \& Parasitic Diseases & 1 & 1 & 0 & 2 \\
\textbf{C} Neoplasms & 4 & 4 & 0 & 1 \\
\textbf{D} Blood Diseases \& Immune Disorders & 5 & 3 & 2 & 2 \\
\textbf{E} Endocrine, Nutritional \& Metabolic Diseases & 9 & 9 & 7 & 6 \\
\textbf{F} Mental \& Behavioural Disorders & 5 & 7 & 3 & 6 \\
\textbf{G} Nervous System Diseases & 1 & 2 & 2 & 6 \\
\textbf{H} Eye, Ear \& Related Diseases & 2 & 5 & 1 & 2 \\
\textbf{I} Circulatory System Diseases & 11 & 9 & 5 & 5 \\
\textbf{J} Respiratory System Diseases & 4 & 7 & 3 & 1 \\
\textbf{K} Digestive System Diseases & 3 & 5 & 5 & 6 \\
\textbf{L} Skin \& Subcutaneous Tissue Diseases & 0 & 0 & 0 & 0 \\
\textbf{M} Musculoskeletal \& Connective Tissue Diseases & 3 & 0 & 5 & 5 \\
\textbf{N} Genitourinary System Diseases & 7 & 6 & 4 & 5 \\
\bottomrule
\end{tabular}%
}
\end{table*}

\clearpage
\section*{Discussion and Conclusion}

Beyond describing cross-nationality differences, the main practical use of comorbidity patterns is to support earlier identification of high-risk multimorbidity profiles and to prioritize targeted prevention or care navigation. Risk-stratified strategies can be evaluated using budget impact analyses, in which earlier detection shifts patients away from costly acute admissions toward lower-intensity management, an approach already demonstrated in systemic lupus erythematosus (M32) \cite{clarke2020evaluation}.

Across nearly all ICD-10 chapters and age groups, Austrian patients show markedly higher prevalence for most conditions (pink), while non-Austrians show elevated rates primarily in obstetric and perinatal categories, a pattern driven partly by the younger demographic profile of the non-Austrian population, Figure \ref{fig:RatioICDs}. Within mental and behavioral disorders, Austrian patients show substantially higher prevalence of substance use, affective, and anxiety-related diagnoses across most age bands, whereas non-Austrians show relative elevations only in selected diagnostic subcategories and age groups, Figure \ref{fig:RatioICDsF}. A similar pattern holds for circulatory diseases, where Austrian patients dominate across the full spectrum of diagnoses, particularly in older age groups, Figure \ref{fig:RatioICDsI}. Musculoskeletal and connective tissue diseases present a notably different picture: here, non-Austrian patients show elevated prevalence ratios across a broad range of diagnoses and midlife age groups in both sexes, consistent with the occupational exposures and physically demanding labor conditions disproportionately experienced by migrant populations, Figure \ref{fig:RatioICDsM}. Together, these figures underscore that the disparities are not simply an artifact of age structure but likely reflect persistent differences in diagnostic depth, occupational risk, and healthcare access.

Both female and male non-Austrians have disproportionately fewer significant comorbidity differences than matched Austrians, Figure \ref{fig:RatioComNet} a. Previously, one study provided evidence that non-Austrians use health care systems differently than Austrians, in that they have lower hospital admission rates but higher readmission rates \cite{dervic2024healthcare}. Lower hospital admission rates, higher readmission rates and significantly fewer comorbidity differences in non-Austrians - regardless of specific nationalities - might be related to variable access barriers within secondary health systems. These can be driven by language barriers, cultural differences in expression of illness complaint, stereotyping or diagnostic bias on the part of medical personnel, as well as a more crisis-oriented hospital admission pattern, all of which might prevent comprehensive diagnostics of comorbidities.
Across ICD-10 chapters and between female and male patients there are numerous differences in patterns of comorbidity between Austrians and non-Austrians, Figure \ref{fig:RatioComNet} a. While in Austrian males comorbidity networks within mental health diagnoses are prominent, this pattern is strikingly smaller in non-Austrian males. Alcohol use disorder (F10) is a highly prevalent condition that is strongly associated with several somatic and psychiatric comorbidities, such as cardiovascular and liver disease, metabolic disorders, infectious diseases and cancer, as well as anxiety and affective disorders as well as other substance use disorders, all together resulting in increased morbidity and premature mortality. Thus, the detection and treatment of alcohol use disorder and its different comorbidities is of relevance for individual and public health. Alcohol consumption in Austria is much above the EU average (12$.$0l vs 11$.$0l APC) with men having a higher consumption than women \cite{who2024alcoholEU, who2024globalstatus}. Our data shows significant differences in the number of comorbidity networks between Austrian and non-Austrian males with alcohol use disorder. For example, comorbidities between alcohol use disorder and several mental health diagnoses, such as depression (F32, F33), reaction to severe stress and adjustment disorder (F43), as well as other substance use disorders (nicotine, F17 and sedatives F13) are significantly more often in Austrian males than in non-Austrian males. In non-Austrian males comorbidities appear significantly more often between alcohol use disorder diagnoses and epileptic seizures (G40) as well as gastro-oesophageal reflux disease (K21). These differences are particularly interesting, because they point towards inequities in time and quality of health care access: while all mentioned comorbidities are typical in combination with alcohol use disorder, epileptic seizures are most commonly seen in the context of severe alcohol withdrawal symptoms, underscoring a more acute presentation of non-Austrian patients, whereas the different F-diagnoses in Austrian males underline the notion of a more comprehensive diagnostic assessment, possibly also in the context of longer hospital stays.

Meanwhile in non-Austrian females comorbidities are prominent within mental health diseases and between these and digestive system as well as musculoskeletal diseases. In both female and male non-Austrians comorbidity networks within musculoskeletal diagnoses are prominent, albeit being less noticeable in Austrians. Furthermore, additional patterns emerged in our analyses when examining comorbidities involving recurrent depression (F33) for females. The co-occurrence of a F33 diagnosis with somatoform disorders (F45) and with dorsalgia (M54) was particularly pronounced among non-Austrian women, while a co-morbid chronic ischemic heart disease (I25) diagnosis was more pronounced among non-Austrian men. These gendered comorbidity patterns may reflect distinct pathways through which the cumulative burden of migration-related (social) risk factors manifests in both psychological and somatic health outcomes. The co-occurrence of recurrent depression (F33) and somatoform disorders (F45) may point to two interrelated phenomena, a more somatoform presentation of underlying depressive syndromes, as well as the existence of structural or linguistic barriers to mental healthcare, which can delay or complicate the accurate diagnosis and adequat treatment of depressive disorders in some migrant groups. The association between recurrent depression (F33) and dorsalgia (M54) in turn, may reflect the overrepresentation of migrants in physically demanding occupations, such as cleaning, as well as an earlier age of entry into the labour market and consequently more prolonged exposure to occupational physical strain. \cite{statistikaustria2025, jestl2026labour, aigner2023left}

\subsubsection*{Strengths and Limitations}

A principal strength of this study is its comprehensive dataset, which includes in-hospital records for about 4 million individuals over 5 years. However, several limitations exist. The dataset lacks information on outpatient care, prescriptions, lifestyle, and patient socioeconomic status. It includes only those who accessed hospital care and received a diagnosis, excluding unmet healthcare needs in the general population. Furthermore, the observed lower hospitalisation rates among non-Austrian patients may partly reflect the healthy migrant effect, whereby migrants tend to be healthier at arrival than the native population, limiting our ability to disentangle this from structural barriers to access. The analysis is also limited to five years and excludes patients with multiple nationalities.

Taken together, our results reveal systematic differences in comorbidity patterns between Austrian and non-Austrian patients that go beyond simple prevalence differences. Non-Austrians have disproportionately fewer significant comorbidity links. They also have lower hospital admission rates and higher readmission rates. These patterns point to inequities in access to and quality of secondary healthcare. This likely does not reflect a lower disease burden. Instead, it is more likely explained by access barriers such as language, cultural factors, and crisis-oriented admission patterns. These barriers may prevent a comprehensive diagnostic assessment of comorbidities. Distinct comorbidity profiles are observed across sex and nationality groups. For example, musculoskeletal and mental-somatic comorbidities are prominent in non-Austrian patients. These findings suggest that migration-related social risk factors shape health outcomes. Current healthcare pathways may not adequately address these risks. Our results underscore the need for targeted, culturally sensitive care strategies for migrant populations. Particular attention should be given to earlier identification of high-risk multimorbidity profiles and improved navigation within the healthcare system.

\clearpage
\section*{Tables}
\setcounter{table}{0}

\begin{table*}[ht]
\caption{General cohort baseline characteristics and selected ICD-10 codes by group}
\centering
\scriptsize
\renewcommand{\arraystretch}{1} 
\setlength{\tabcolsep}{3pt} 
\begin{tabular}{c|lll|lll}
& \multicolumn{3}{c|}{\textbf{Austrians}} 
 & \multicolumn{3}{c}{\textbf{Non-Austrians}} \\
 & All & Female & Male & All & Female & Male  \\ 
    \hline
        \hline
        All patients & 3,607,058 & 1,927,211 & 1,679,847 & 327,394 & 189,653 & 137,741 \\
Age & 53$.$32$\pm$24$.$18 & 53$.$98$\pm$24$.$06 & 52$.$56$\pm$24$.$3 & 39$.$81$\pm$18$.$84 & 39$.$33$\pm$17$.$34 & 40$.$45$\pm$20$.$72 \\
Hospital\_stay & 2$.$87$\pm$4$.$16 & 2$.$83$\pm$4$.$02 & 2$.$91$\pm$4$.$33 & 2$.$09$\pm$3$.$34 & 2$.$06$\pm$2$.$94 & 2$.$14$\pm$3$.$81 \\
Hospital\_days & 16$.$07$\pm$31$.$56 & 15$.$89$\pm$31$.$31 & 16$.$29$\pm$31$.$85 & 8$.$81$\pm$21$.$83 & 8$.$27$\pm$19$.$36 & 9$.$56$\pm$24$.$81 \\
Number\_diagnoses & 2$.$24$\pm$1$.$92 & 2$.$22$\pm$1$.$88 & 2$.$25$\pm$1$.$97 & 1$.$67$\pm$1$.$25 & 1$.$69$\pm$1$.$21 & 1$.$65$\pm$1$.$31 \\

E11 & 6$.$57 & 5$.$54 & 7$.$76 & 3$.$86 & 2$.$86 & 5$.$23 \\
E66 & 3$.$82 & 3$.$7 & 3$.$96 & 3$.$19 & 3$.$24 & 3$.$12 \\
F32 & 3$.$21 & 4 & 2$.$3 & 1$.$78 & 1$.$92 & 1$.$59 \\
F33 & 1$.$6 & 1$.$95 & 1$.$21 & 0$.$96 & 0$.$99 & 0$.$93 \\
I10 & 21$.$76 & 20$.$83 & 22$.$83 & 9$.$38 & 7$.$43 & 12$.$06 \\
I21 & 1$.$83 & 1$.$25 & 2$.$49 & 0$.$97 & 0$.$4 & 1$.$76 \\
I50 & 4$.$3 & 4$.$11 & 4$.$53 & 1$.$07 & 0$.$82 & 1$.$42 \\
I63 & 2$.$27 & 2$.$02 & 2$.$55 & 0$.$71 & 0$.$47 & 1$.$05 \\
J18 & 3$.$72 & 3$.$16 & 4$.$37 & 1$.$73 & 1$.$24 & 2$.$41 \\
C00-C99 & 9$.$14 & 8$.$03 & 10$.$41 & 4$.$31 & 3$.$7 & 5$.$16 \\ 
E00-E99 & 22$.$92 & 22$.$65 & 23$.$23 & 13$.$3 & 12$.$06 & 15 \\ 
F00-F99 & 13$.$17 & 13$.$17 & 13$.$18 & 9$.$55 & 7$.$68 & 12$.$13 \\ 
I00-I99 & 32$.$32 & 30$.$31 & 34$.$63 & 15$.$56 & 12$.$24 & 20$.$13 \\ 
O00-O99 & 7$.$22 & 13$.$51 & 0 & 22$.$21 & 38$.$34 & 0 \\ 
   \hline
\hline
\end{tabular}
\end{table*}

\clearpage

\begin{table*}[ht]
\caption{Matched cohort baseline characteristics and selected ICD-10 codes by group}
\centering
\scriptsize
\renewcommand{\arraystretch}{1} 
\setlength{\tabcolsep}{3pt} 
\begin{tabular}{c|lll|lll}
& \multicolumn{3}{c|}{\textbf{Matched Austrians}} 
 & \multicolumn{3}{c}{\textbf{Non-Austrians}} \\
 & All & Female & Male & All & Female & Male  \\ 
    \hline
        \hline
        All patients & 272,779 & 159,280 & 113,499 & 272,779 & 159,280 & 113,499 \\
Age & 40$.$04$\pm$19$.$06 & 39$.$52$\pm$17$.$52 & 40$.$78$\pm$21 & 39$.$94$\pm$18$.$85 & 39$.$43$\pm$17$.$31 & 40$.$66$\pm$20$.$81 \\
Hospital stay & 2$.$38$\pm$3$.$49 & 2$.$35$\pm$3$.$38 & 2$.$43$\pm$3$.$63 & 2$.$26$\pm$3$.$6 & 2$.$21$\pm$3$.$16 & 2$.$32$\pm$4$.$14 \\
Hospital days & 11$.$35$\pm$27$.$3 & 10$.$8$\pm$26$.$7 & 12$.$11$\pm$28$.$11 & 9$.$89$\pm$23$.$42 & 9$.$23$\pm$20$.$7 & 10$.$81$\pm$26$.$75 \\
Number diagnoses & 1$.$91$\pm$1$.$52 & 1$.$91$\pm$1$.$46 & 1$.$92$\pm$1$.$6 & 1$.$77$\pm$1$.$33 & 1$.$78$\pm$1$.$28 & 1$.$76$\pm$1$.$4 \\

E11 & 2$.$84 & 1$.$8 & 4$.$29 & 4$.$18 & 3$.$09 & 5$.$71 \\
E66 & 3$.$2 & 2$.$98 & 3$.$52 & 3$.$45 & 3$.$52 & 3$.$34 \\
F32 & 2$.$32 & 2$.$57 & 1$.$96 & 1$.$97 & 2$.$12 & 1$.$76 \\
F33 & 1$.$46 & 1$.$59 & 1$.$27 & 1$.$06 & 1$.$09 & 1$.$03 \\
I10 & 9$.$38 & 7$.$03 & 12$.$68 & 10$.$05 & 7$.$92 & 13$.$03 \\
I21 & 0$.$85 & 0$.$37 & 1$.$52 & 1$.$03 & 0$.$43 & 1$.$88 \\
I50 & 1$.$17 & 0$.$79 & 1$.$71 & 1$.$18 & 0$.$91 & 1$.$57 \\
I63 & 0$.$87 & 0$.$61 & 1$.$24 & 0$.$76 & 0$.$51 & 1$.$12 \\
J18 & 1$.$68 & 1$.$22 & 2$.$32 & 1$.$9 & 1$.$35 & 2$.$67 \\
C00-C99 & 4$.$92 & 4$.$18 & 5$.$96 & 4$.$5 & 3$.$81 & 5$.$48 \\ 
E00-E99 & 13$.$89 & 12$.$57 & 15$.$76 & 14$.$11 & 12$.$71 & 16$.$06 \\ 
F00-F99 & 10$.$66 & 9$.$54 & 12$.$22 & 10$.$17 & 8$.$19 & 12$.$96 \\ 
I00-I99 & 16$.$88 & 13$.$5 & 21$.$62 & 16$.$4 & 12$.$9 & 21$.$31 \\ 
O00-O99 & 17$.$64 & 30$.$21 & 0 & 22$.$96 & 39$.$32 & 0 \\ 
\hline
\end{tabular}
\end{table*}

\clearpage

\begin{table*}[!h]
\centering
\scriptsize
\caption{Distribution of significant comorbidity differences across ICD-10 chapters by gender and nationality}
\resizebox{\textwidth}{!}{%
\footnotesize
\begin{tabular}{lcccc}
\toprule
 & \multicolumn{2}{c}{Austrians} & \multicolumn{2}{c}{Non-Austrians} \\
\cmidrule(lr){2-3} \cmidrule(lr){4-5}
Chapter & Female & Male & Female & Male \\
\midrule
\textbf{A} Infectious \& Parasitic Diseases & 2 & 2 & 0 & 1 \\
\textbf{B} Infectious \& Parasitic Diseases & 1 & 1 & 0 & 2 \\
\textbf{C} Neoplasms & 4 & 4 & 0 & 1 \\
\textbf{D} Blood Diseases \& Immune Disorders & 5 & 3 & 2 & 2 \\
\textbf{E} Endocrine, Nutritional \& Metabolic Diseases & 9 & 9 & 7 & 6 \\
\textbf{F} Mental \& Behavioural Disorders & 5 & 7 & 3 & 6 \\
\textbf{G} Nervous System Diseases & 1 & 2 & 2 & 6 \\
\textbf{H} Eye, Ear \& Related Diseases & 2 & 5 & 1 & 2 \\
\textbf{I} Circulatory System Diseases & 11 & 9 & 5 & 5 \\
\textbf{J} Respiratory System Diseases & 4 & 7 & 3 & 1 \\
\textbf{K} Digestive System Diseases & 3 & 5 & 5 & 6 \\
\textbf{L} Skin \& Subcutaneous Tissue Diseases & 0 & 0 & 0 & 0 \\
\textbf{M} Musculoskeletal \& Connective Tissue Diseases & 3 & 0 & 5 & 5 \\
\textbf{N} Genitourinary System Diseases & 7 & 6 & 4 & 5 \\
\bottomrule
\end{tabular}%
}
\end{table*}

\clearpage

\section*{Figures} 
\setcounter{figure}{0}

\begin{figure*}[!h]
\centering
{\includegraphics[width=0.99\linewidth]{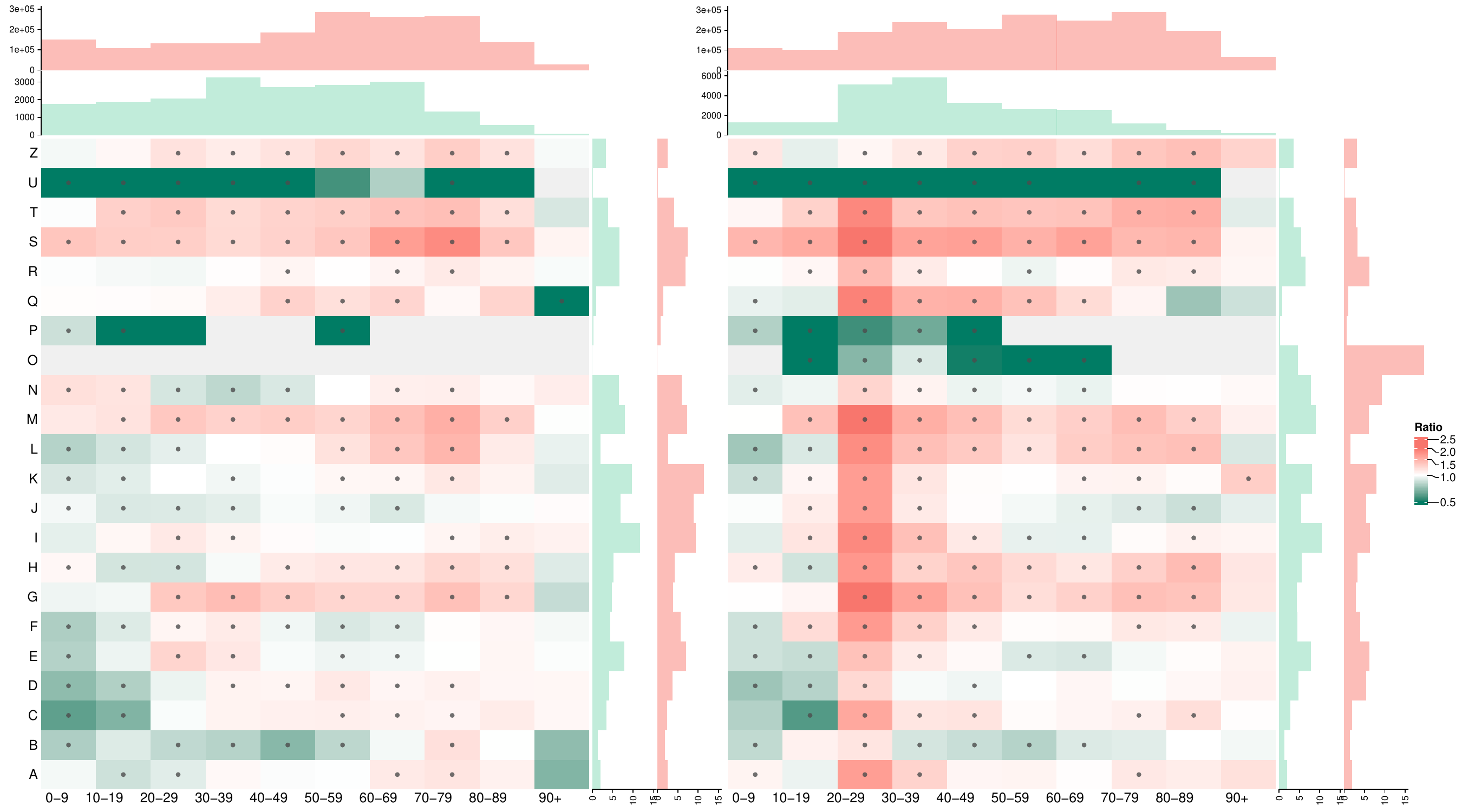}}
\caption{Log-scale prevalence ratios across ICD-10 chapters comparing non-Austrian and Austrian patients, stratified by sex (females: right; males: left). Colored squares represent prevalence differences, with green indicating higher prevalence among non-Austrians and pink among Austrians; grey dots mark statistically significant differences ($p < 0.05$). Grey squares denote cohorts too small to estimate prevalence ratios. Marginal histograms display patient distributions by age group (top) and ICD-10 chapter (right; note: scales differ).}
\end{figure*}

\clearpage

\begin{figure*}[!h]
\centering
{\includegraphics[width=0.90\linewidth]{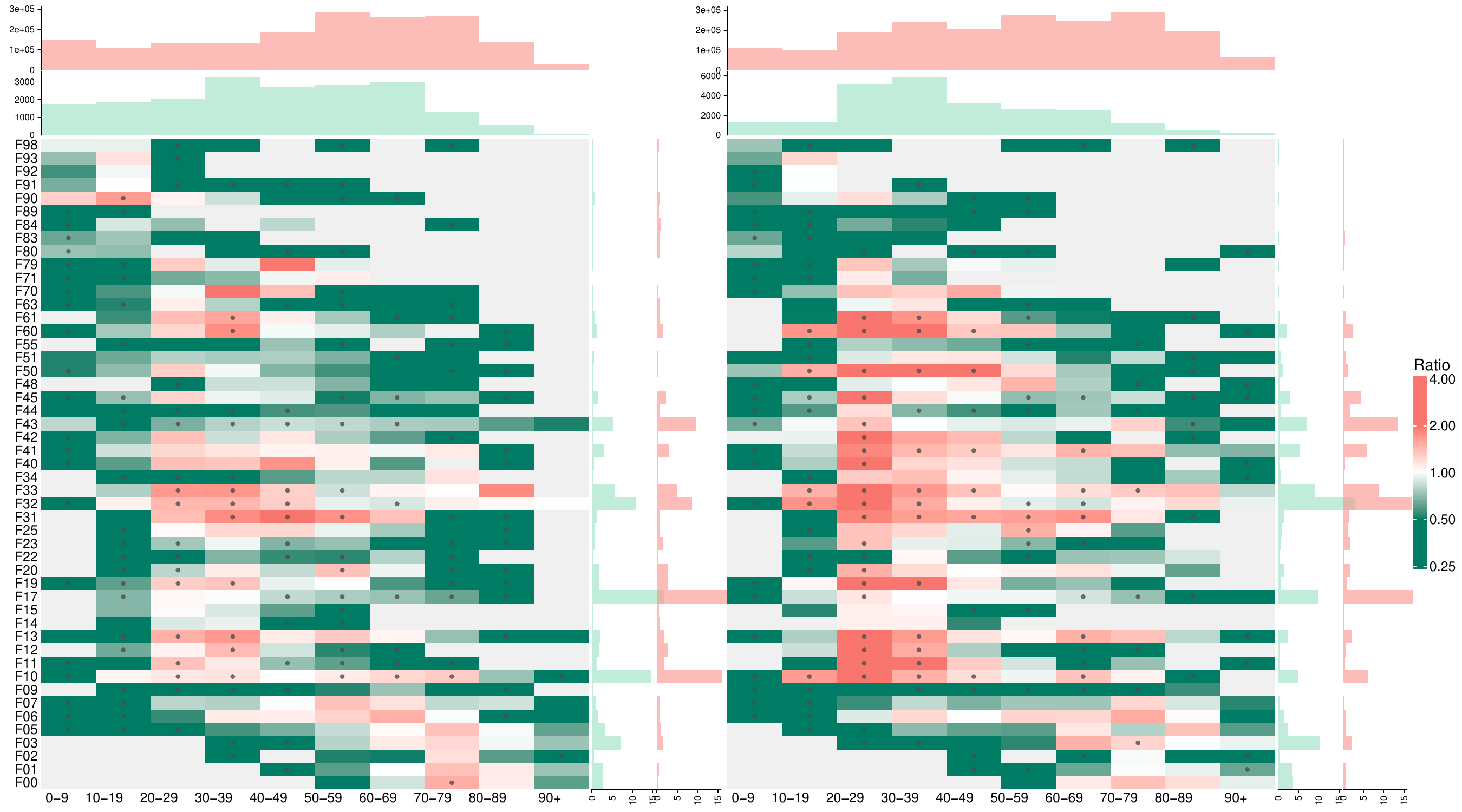}}
\caption{Log-scale ratios of disease prevalence across ICD-10 chapter \textbf{F} - Mental \& behavioural disorders - comparing non-Austrian and Austrian patients, stratified by sex (females: right; males: left). Colored squares represent prevalence differences, with green indicating higher prevalence among non-Austrians and pink among Austrians; grey dots mark statistically significant differences ($p < 0.05$). Grey squares denote cohorts too small to estimate prevalence ratios. Marginal histograms display patient distributions by age group (top) and ICD-10 chapter (right; note: scales differ).}

\end{figure*}

\clearpage
\begin{figure*}[!h]
\centering
{\includegraphics[width=0.90\linewidth]{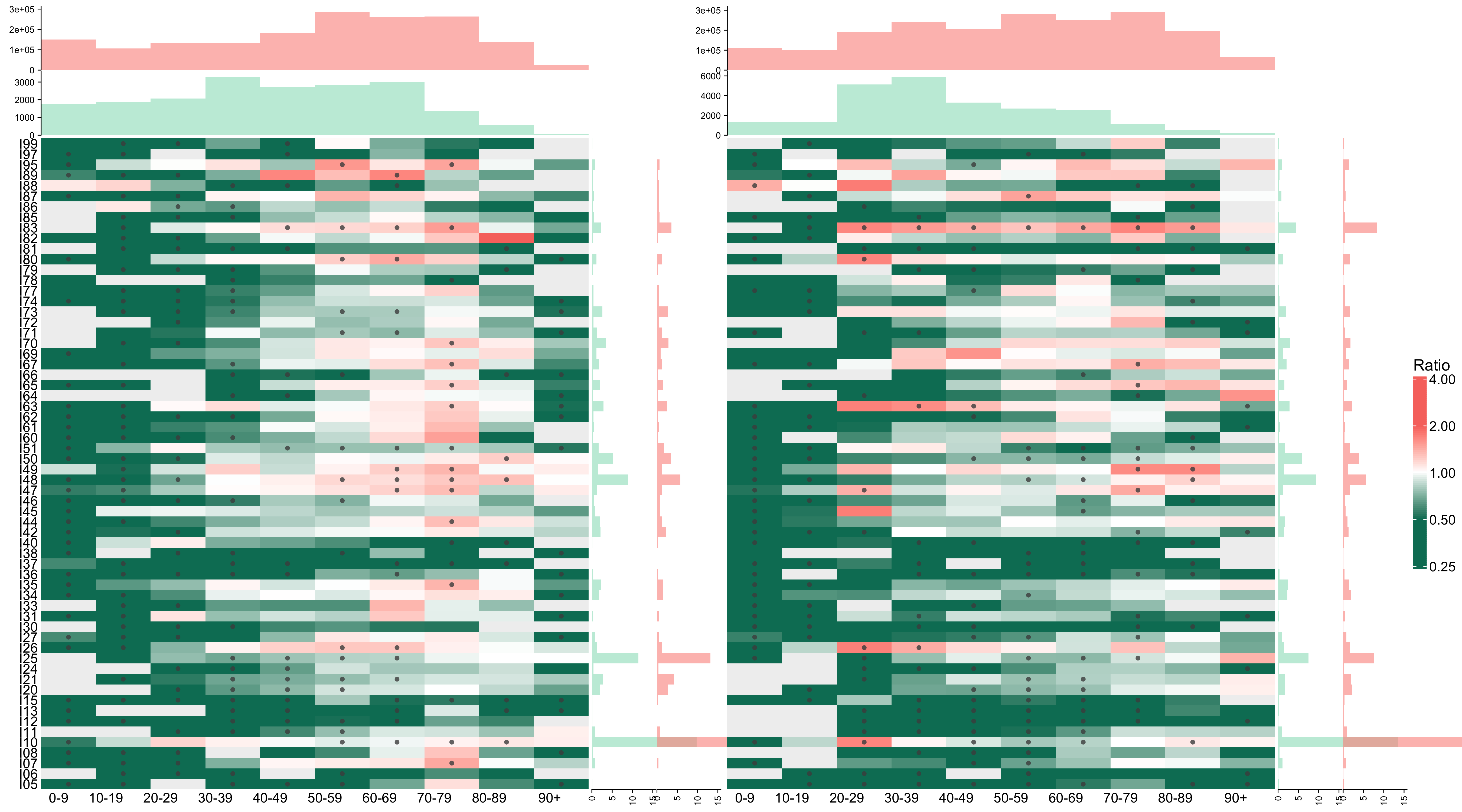}}
\caption{Log-scale ratios of disease prevalence across ICD-10 chapter \textbf{I} - Circulatory system diseases - comparing non-Austrian and Austrian patients, stratified by sex (females: right; males: left). Colored squares represent prevalence differences, with green indicating higher prevalence among non-Austrians and pink among Austrians; grey dots mark statistically significant differences ($p < 0.05$). Grey squares denote cohorts too small to estimate prevalence ratios. Marginal histograms display patient distributions by age group (top) and ICD-10 chapter (right; note: scales differ).}
\end{figure*}

\clearpage

\begin{figure*}[!h]
\centering
{\includegraphics[width=0.90\linewidth]{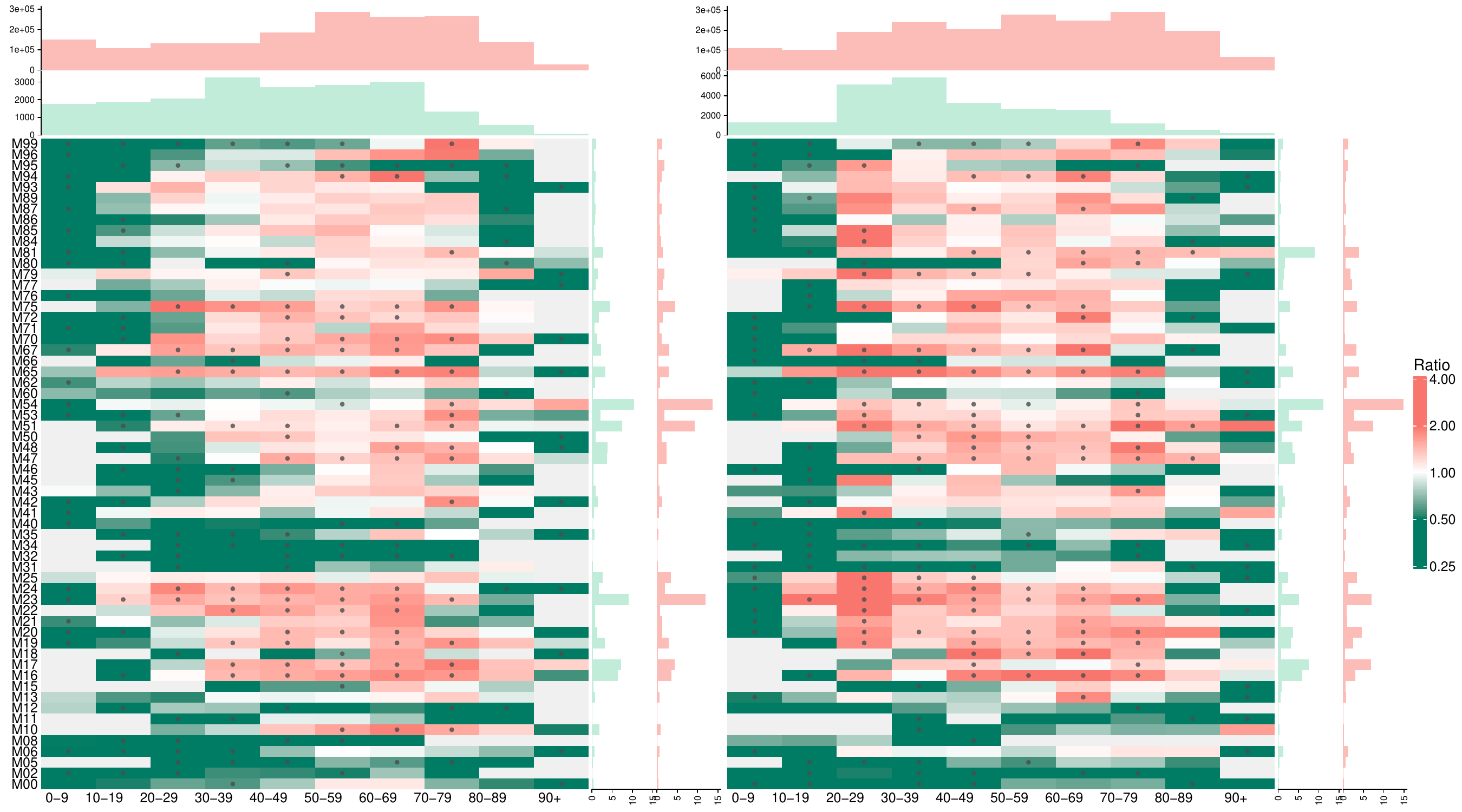}}
\caption{Log-scale ratios of disease prevalence across ICD-10 chapter \textbf{M} - Diseases of the musculoskeletal system and connective tissue - comparing non-Austrian and Austrian patients, stratified by sex (females: right; males: left). Colored squares represent prevalence differences, with green indicating higher prevalence among non-Austrians and pink among Austrians; grey dots mark statistically significant differences ($p < 0.05$). Grey squares denote cohorts too small to estimate prevalence ratios. Marginal histograms display patient distributions by age group (top) and ICD-10 chapter (right; note: scales differ).}
\end{figure*}

\clearpage

\begin{figure*}[!h]
\centering
{\includegraphics[width=0.99\linewidth]{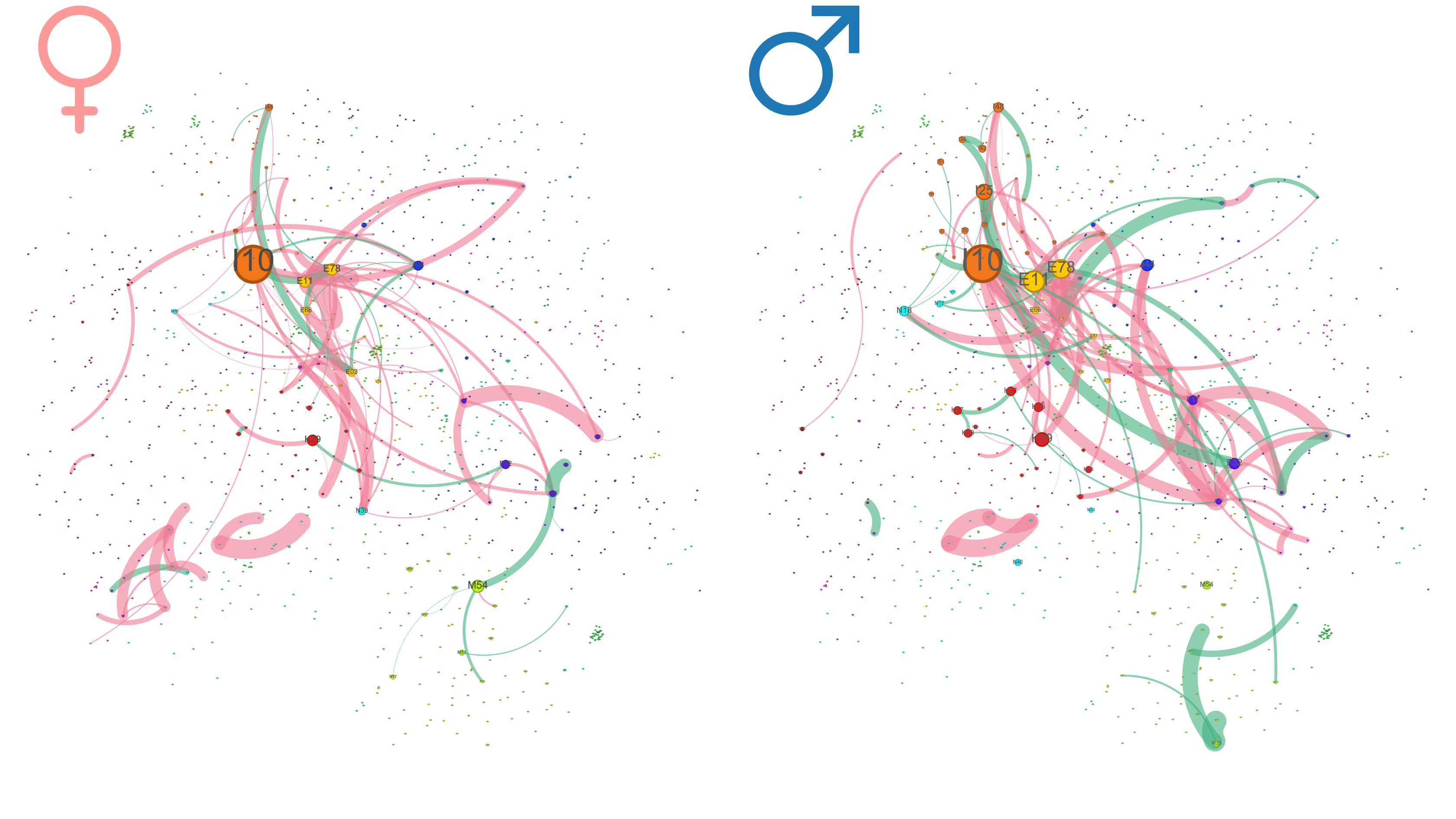}}

\caption{Comorbidity networks highlighting diagnoses that differ significantly between Austrian and non-Austrian patients. Nodes represent ICD-10 3-digit diagnoses; their size reflects node degree, and color denotes the corresponding ICD chapter (based on the first ICD-10 letter). Links represent comorbidities with statistically significant differences between the two groups. Link weights are proportional to the difference in log odds ratios (ORs), scaled by the pooled standard error. Link color indicated the direction of the difference, with pink links denoting stronger co-occurrence among Austrians and green links denoting stronger co-occurrence among non-Austrians. An interactive version is 
available at \url{https://vis.csh.ac.at/diaspora_model_for_migration/migration-net/}} 
\end{figure*}

\clearpage

\begin{figure*}[!ht]
\centering
{\includegraphics[width=0.99\linewidth]{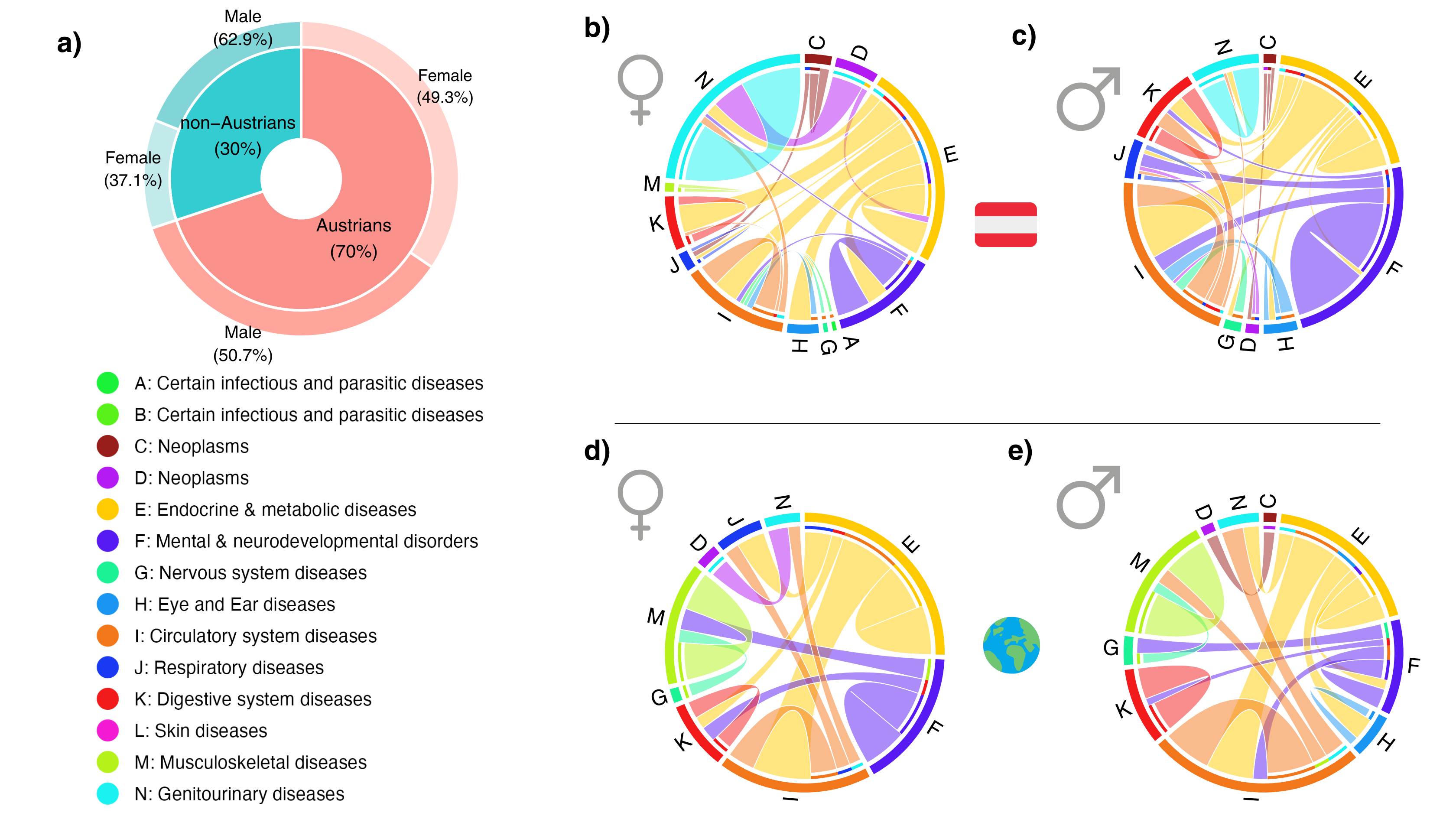}}
\caption{Distribution of significant comorbidity differences a) nationality b) across ICD-10 chapters for males Austrians, c) females Austrians, d) males non-Austrians e) females non-Austrians} 
\end{figure*}

\clearpage

 \section*{Code Availability}
Custom code for the analysis is available upon request from the authors.

\section*{Competing Interests}
The authors declare no competing interests.

\section*{Author Contributions}
ED conceived the study, carried out the analysis, and produced the plots and graphics. AV and RD conducted the literature review. MA and NM contributed medical expertise in interpreting the findings. All authors contributed to writing and reviewing the manuscript.


\renewcommand{\baselinestretch}{1.5}


\renewcommand{\baselinestretch}{1.5}

\bibliographystyle{naturemag} 
\bibliography{references}     

\end{document}